\documentclass[
reprint,
 amsmath,amssymb,
 aps,
prl,
]{revtex4-2}

\usepackage{graphicx}
\usepackage{dcolumn}
\usepackage{bm}
\usepackage{xcolor}
\usepackage{amsmath}
\usepackage{float} 
\usepackage{physics}
\usepackage{url}
\usepackage{chemfig}
\usepackage[colorinlistoftodos]{todonotes}

\begin{document}

\preprint{APS/123-QED}

\title{Stopping cross-section for protons across different phases of water}






\author{F. Matias\textsuperscript{1}}
\author{N. E. Koval\textsuperscript{2}}
\author{P. de Vera\textsuperscript{3}}
\author{R. Garcia-Molina\textsuperscript{3}}
\author{I. Abril\textsuperscript{4}}
\author{J. M. B. Shorto\textsuperscript{1}}
\author{H. Yoriyaz\textsuperscript{1}}
\author{J. J. N. Pereira\textsuperscript{1}}
\author{T. F. Silva\textsuperscript{5}}
\author{M. H. Tabacniks\textsuperscript{5}}
\author{M. Vos\textsuperscript{6}}


\author{P. L. Grande\textsuperscript{7}}

\affiliation{\textsuperscript{1}\mbox{Instituto de Pesquisas Energéticas e Nucleares, Av. Professor Lineu Prestes, São Paulo, 05508-000, Brazil}}

\affiliation{\textsuperscript{2}\mbox{Centro de F\'isica de Materiales, Paseo Manuel de Lardiz\'abal 5, Donostia-San Sebasti\'an, 20018, Spain}}

\affiliation{\textsuperscript{3}\mbox{Departamento de Física, Centro de Investigación en Óptica y Nanofísica, Universidad de Murcia, Murcia, Spain}}

\affiliation{\textsuperscript{4}\mbox{Departament de Física Aplicada, Universitat d’Alacant, Alacant, Spain}}

\affiliation{\textsuperscript{5}\mbox{Instituto de Física da Universidade de São Paulo, Rua do Matão, trav. R187, São Paulo, 05508-090, Brazil}}

\affiliation{\textsuperscript{6}\mbox{Department of Materials Physics, Research School of Physics, Australian National University (ANU), Canberra, Australia}}

\affiliation{\textsuperscript{7}\mbox{Instituto de Física da Universidade Federal do Rio Grande do Sul, Av. Bento Gonçalves, Porto Alegre, 9500, Brazil}}



\date{\today}

\begin{abstract}

\noindent
Accurately quantifying the energy loss rate of proton beams in liquid water is crucial for the precise application and improvement of proton therapy, whereas the slowing down of protons in water ices also plays an important role in astrophysics.
However, 
precisely determining the electronic stopping power, particularly for the liquid phase, 
has been elusive so far. 
Experimental techniques are difficult to apply to volatile liquids, and the availability of sufficient reliable 
measurements has been limited to
the solid and vapor phases. 
The accuracy of current models is typically limited to proton energies just above the energy-loss maximum, making it difficult to predict radiation effects at an energy range of special relevance. We elucidate the phase differences in proton energy loss in water in a wide energy range ($0.001 - 10$ MeV) 
by means of real-time time-dependent density functional theory combined with the Penn method. 
This non-perturbative model, more computationally-efficient than current approaches, 
describes the phase effects in water 
in excellent agreement with available experimental data, revealing 
clear deviations around the maximum of the stopping power curve and below. As an important outcome, our calculations reveal that proton stopping quantities of liquid water and amorphous ice are 
identical, 
in agreement with recent similar observations for low-energy electrons, pointing out to 
this equivalence 
for all charged particles. This could help to overcome the limitation in obtaining reliable experimental information for the biologically-relevant liquid water target.
%

\end{abstract}

\maketitle



Proton therapy is one of the most advanced tools for cancer treatment \cite{Durante_2016, Schardt2010}, as it offers more precision in energy delivery to tumor zones and less damage to healthy tissues than traditional radiotherapy, owing to its characteristic depth-dose profile. Known as the Bragg peak, this profile maximizes the energy delivery by the end of the protons trajectories while sparing the surrounding areas. Understanding how energetic protons lose their energy in biological matter 
is crucial for the precise control of the radiation dose, importantly affecting the sub-milimetric positioning of the Bragg peak within the patient \cite{Andreo2009,Paul2013ACQ} needed for treatment planning.
Liquid water is typically used as a target when studying radiation effects 
due to its biological relevance \cite{Nikjoo2012,Nikjoo_2016},
making it an ideal material for effectively investigating the fundamentals of biodamage. 
The rate at which protons lose energy in liquid water sets the initial conditions for the molecular mechanisms responsible for radiation damage, related to DNA clustered lesions directly produced by secondary electron impact, or indirectly by free radical production \cite{Nikjoo_2016}. Uncertainties around the energy-loss maximum (low proton energies, $\leq 0.2$ MeV) limit the predictive power of biophysical models for proton effects in tissue.

In addition to its relevance for proton therapy and the fundamental understanding of proton-matter interaction, studying the different phases of water is also important in astrophysics, where the interaction of cosmic energetic protons with ices plays a significant role
\cite{BARAGIOLA2003953,farenzena2005}. Water ice is abundant on comets, interstellar dust and planetary moon's surfaces \cite{doi:10.1126/science.1258055,Dartois2024,Protopapa2024,doi:10.1126/science.1123632} which are constantly bombarded by solar wind primarily consisting of protons and electrons. Proton irradiation of ice can lead to radiolysis and formation of reactive species including radicals \cite{https://doi.org/10.1002/2017JE005285} as well as to amorphization of crystalline ice \cite{BARAGIOLA2005187}. The radiolysis products are crucial for understanding chemical processes in extraterrestrial environments. In astrochemistry, they help explain the chemical evolution of icy bodies in space \cite{Jenniskens1994,Mejia2024} contributing to the synthesis of organic macromolecules \cite{Ligterink2024}, and may play a role in the chemical enrichment of protoplanetary disks. Understanding proton energy loss in various water phases is therefore essential, as it determines the depth and efficiency of energy deposition that drives such processes.

Despite its importance as radiation target both in biomedical applications and astrophysics, the precise values of the electronic stopping power (ESP, average energy loss per unit path length of the proton) of water around its maximum (proton energies $\leq 0.2$ MeV) are uncertain, particularly for the liquid phase. Volatile liquids such as water are difficult to manipulate in energy loss experiments and, thus, the empirical information on its ESP is scarce \cite{SHIMIZU20092667,SHIMIZU20101002,Siiskonen2011}. The interpretation of experimental measurements closer to the stopping maximum is based on Monte Carlo simulations \cite{SHIMIZU20092667,SHIMIZU20101002}, but the accuracy of the extracted values has been under debate \cite{GARCIAMOLINA201351,deVera2019jets}. Measurements are more straightforward with solid and gas water targets, for which experimental information is more abundant \cite{BAUER1994132,PhysRev.90.532,PhysRev.92.742,MITTERSCHIFFTHALER199058,baek2006}. Even though there have been recent claims that liquid water slows down charged particles (specifically low-energy electron beams) similarly to amorphous ice \cite{Signorell2020}, this discussion is not yet settled, still less for proton beams.

The issue is also problematic from the theoretical point of view. State-of-the-art calculations for the proton beam ESP in condensed matter are based mainly on the dielectric formalism \cite{penn1987,Nikjoo2012,GarciaMolina2012SpringerScience,Maarten:2019b,deVera2023} or on first-principles methods such as time-dependent density functional theory (TDDFT) \cite{Reeves2016,kanai:2019,gu2020,taioli2021,koval2022,GU2022109961,doi:10.1021/acs.jpcb.3c05446,koval2023,Shepard2023}. The former methodology, even though it can accurately account for the target phase state (by means of the material's electronic excitation spectrum via its energy loss function, ELF), presents lower accuracy at energies at and below the stopping maximum, due to its perturbative nature. This is precisely the energy range at which the phase effects are expected to be more important. The latter approaches, when applied to atomistic structures, are notably accurate but can be computationally demanding, mainly because of the number of trajectories required to calculate the average ESP for random orientations of the target. There is thus the need for reliable, yet computationally-efficient, non-perturbative theoretical approaches that can provide accurate ESP results even below the stopping maximum, and which can clarify the influence of phase state in the proton energy loss in water.


To overcome these limitations, in this work, we use a more efficient computational procedure, which combines real-time TDDFT \cite{borisov2004,quijada2007,koval2013,matias2017} and the Penn method \cite{penn1987,Maarten:2019b}, which we 
refer to as the TDDFT-Penn approach \cite{matias:2024}. Instead of simulating an explicit atomistic structure, the interaction of an energetic proton with a nearly homogeneous electron gas (jellium) of a given density is efficiently treated by real-time TDDFT. Then, the Penn model is used to represent the inhomogeneities of the target's electron density by means of its ELF. As previously shown for such complex targets as polymers \cite{matias:2024}, the non-perturbative character of the TDDFT method allows the precise calculation of proton ESP at energies even below the ESP maximum, while the Penn approach realistically reflects the particular electronic structure of the complex material. We will show that this method can accurately reproduce ESP measurements for the different phases, showing that liquid water and amorphous ice slow protons down at the same rate. This points out the practical use of amorphous ice targets to experimentally determine the elusive ESP of liquid water.

A detailed description of the TDDFT-Penn methodology is given in the  Supplemental Material \cite{suppmat}  (see also references \cite{PhysRevB.13.4274,PhysRevLett.81.4831,Koval2017}). In brief, the water media in different phases is modeled as a collection of jellium spheres of uniform positive background density $n_\mathrm{0}(r_s)$, defined by the Wigner-Seitz radius $r_s$, through the relation $4\pi r_{s}^{3}/3=1/n_\mathrm{0}$. The ESP of a jellium sphere for a proton of velocity $v$ crossing through its center, $\left[{\rm d}E/{\rm d}z(v)\right]_{\text{TDDFT}}$, can be straightforwardly calculated by real-time TDDFT in the Kohn-Sham regime, integrating the time-dependent induced force along the proton's trajectory. 
This methodology has been successfully applied to calculate the ESP of various metallic and nonmetallic targets \cite{PhysRevLett.81.4831,quijada2007,Koval2017}. The jellium density is related to its plasmon resonance frequency $\omega_\mathrm{p}$ which, in the optical limit, is given by $\omega_\mathrm{p}^2=4\pi n_\mathrm{0}$, so the ESP for a given plasmon frequency is labeled as $\left[{\rm d}E/{\rm d}z(v,\omega_\mathrm{p})\right]_{\text{TDDFT}}$.
The inhomogeneous electron density of the real target material is introduced by a weighted sum over the ESP of jelliums:
\begin{equation}
\begin{split}
        \left[\frac{{\rm d}E}{{\rm d}z}(v)\right]_{\text{TDDFT-Penn}}\\ 
        = \int_0^{\infty}{\rm d}\omega_\mathrm{p} \, g(\omega_\mathrm{p})
        \left[\frac{{\rm d}E}{{\rm d}z}(v,\omega_\mathrm{p})\right]_{\text{TDDFT}}.
    \label{eq:penn}
\end{split}
\end{equation}
As each jellium is identified by its plasmon frequency in the optical limit $\omega_\mathrm{p}$, the weighting function $g(\omega_\mathrm{p})$ can be obtained from the material's optical ELF, according to the Penn model \cite{penn1987,Maarten:2019b}:
\begin{eqnarray}
       g(\omega_\mathrm{p}) =\frac{2}{\pi \omega_\mathrm{p}}\text{ELF}(\omega_\mathrm{p}).
    \label{eq:elf} 
\end{eqnarray}
\noindent

In our approach, the optical ELF is used to define a physically motivated statistical distribution of local plasmon frequencies within the Penn model. This weighting distribution is then coupled with fully non-perturbative real-time TDDFT simulations of the proton traversing a jellium sphere, which dynamically capture both the energy and momentum transfer processes. This hybrid scheme combines the advantages of macroscopic dielectric models with a microscopic time-dependent treatment of the electron response. The use of the optical ELF in this context is physically meaningful, as it encapsulates the essential spectral characteristics of the material, ensuring the fulfillment of the $f$-sum rule and providing the correct mean excitation energy $I_{\rm Bethe}$, which dominates the ESP at high proton velocities. 
This procedure has been shown to correctly reproduce both the low- and the high-energy stopping regimes, converging to the Bethe limit, as discussed in \cite{Maarten:2019b}.

\begin{figure}[H]
    \centering
\includegraphics[width=0.48\textwidth]{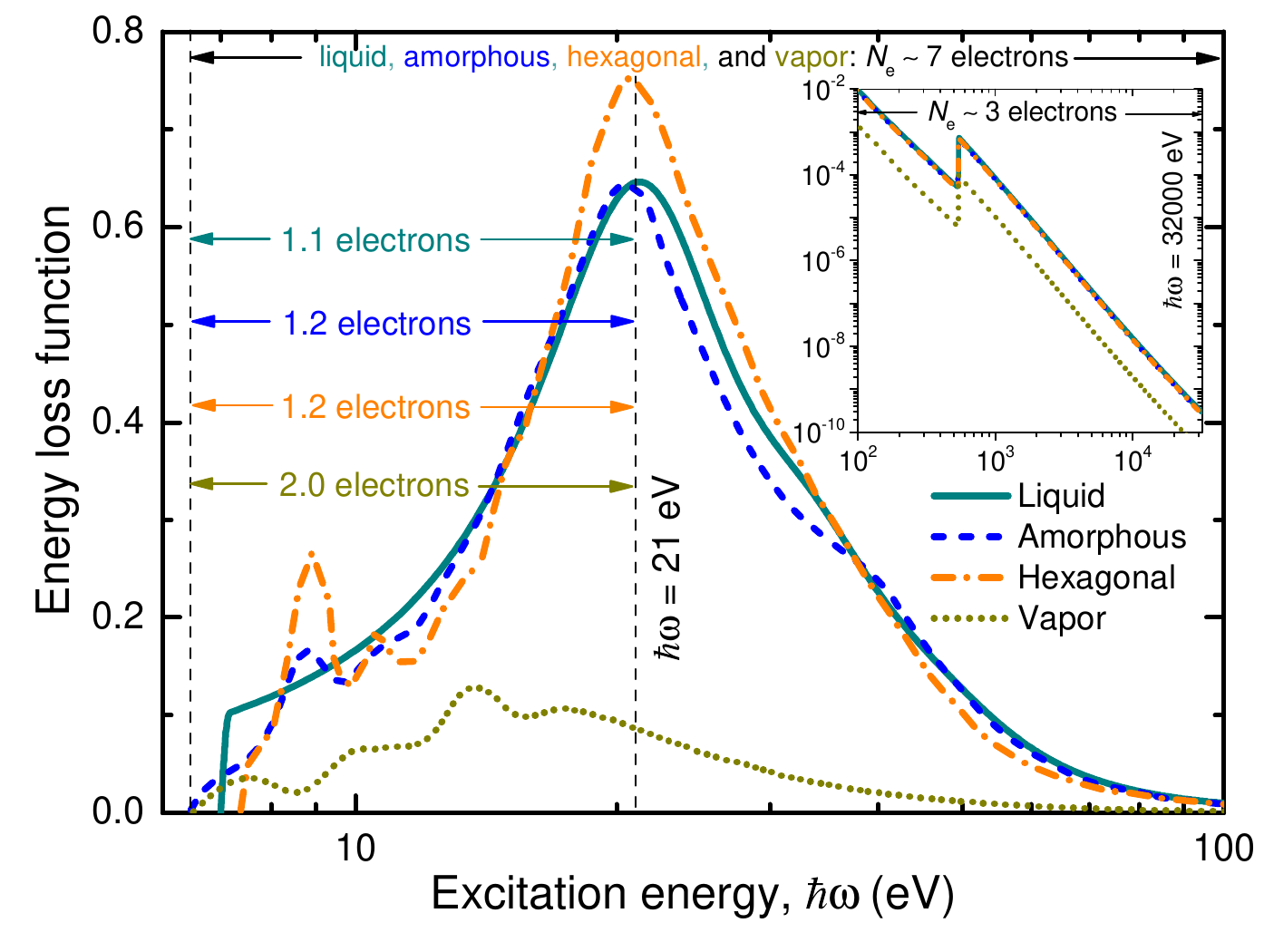}
    \caption{Energy-loss functions for liquid water \cite{Hayashi2000},
    amorphous \cite{DANIELS1971240} 
    and hexagonal ices \cite{Kobayashi1983},
    and water vapor \cite{CHAN1993387} in the optical limit ($\hbar k=0$) as a function of the transferred energy, $\hbar \omega$. The curves result from fitting experimental data ($\hbar \omega < 100$ eV) of liquid water \cite{GARCIAMOLINA20092647}, amorphous ice \cite{EMFIETZOGLOU2007141}, hexagonal ice \cite{EMFIETZOGLOU2007141}, and water vapor. 
    }
    \label{fig:elfs}
\end{figure}

The optical ELF of the different water phases (liquid \cite{Hayashi2000},
amorphous \cite{DANIELS1971240}
and hexagonal ices \cite{Kobayashi1983},
and vapor \cite{CHAN1993387}) are shown in Figure~\ref{fig:elfs} as a function of the excitation energy $\hbar \omega$, where the similarities and differences (particularly for the vapor phase) due to the outer electrons are evident, indicating different electronic structures.
The shifts on the position of the maxima are a well-known consequence of molecular aggregation \cite{Vos2025}.
The contribution to the ELF due to the inner-shell electrons, depicted in the inset of Figure~\ref{fig:elfs}, shows the same structure (except from scale) as it is due to electrons not participating in the specific bonds of each phase.
The validity of these ELFs is assessed by the fulfillment of the $f$-sum rule \cite{Smith1998}, which in the range 0–32000 eV provides the total number of electrons of a water molecule for each phase, as indicated in Figure~\ref{fig:elfs} (namely, $\sim 7$ electrons for $\hbar \omega \leq 100$ eV and $\sim 3$ electrons for $\hbar \omega > 100$ eV). When applying the $f$-sum rule up to $21$ eV, the vapor phase gives approximately twice as many electrons per molecule as the others, which can be taken as an indication of reduced screening effects with respect to the condensed phases, affecting proton energy loss. It should be noted that the ELF is a macroscopic quantity, the intensity of which is proportional to the target molecular density, whereas the effective numbers of electrons are calculated per molecule, so the outcome of the $f$-sum rule is not determined by the absolute differences in ELF intensity between phases, but rather by the relative positions and intensities of the different peaks.

Real-time TDDFT-Penn calculations of the ESP were performed from Eqs.~(\ref{eq:penn}) and~(\ref{eq:elf}), 
as explained in ref.~\cite{matias:2024} and in the Supplemental Material \cite{suppmat}.
Based on the experimental data of the water phases, the corresponding ELF is obtained analytically in the optical limit (null momentum transfer, $\hbar k=0$) employing the Mermin energy loss function-generalized oscillator strengths (MELF-GOS) methodology \cite{HerediaAvalos2005a,GarciaMolina2012SpringerScience,deVera2023}, where the main features of the optical ELF for outer electron excitations ($\hbar \omega \lesssim 100$ eV) are fitted with Mermin-type energy loss functions (MELF) \cite{Mermin1970}. The high-energy part of the ELF, associated with the excitation of atomic inner-shell electrons, is described by hydrogenic generalized oscillator strengths (GOS) \cite{Egerton2011}. Although the MELF-GOS method provides a scheme for the dispersion of the ELF over $k>0$ starting from the optical limit \cite{HerediaAvalos2005a,GarciaMolina2012SpringerScience}, such a dispersion is not needed for the present calculations, as the momentum transfers are implicitly accounted for in the TDDFT simulations.

In order to assess the effect of water phase on the ESP due to electronic effects only (excluding the obvious density differences), it is convenient to analyze the stopping cross section (SCS), namely the ESP divided by the mass density of the target ($1$ g/cm$^3$ for liquid water, $0.94$ g/cm$^3$ \cite{Jenniskens1994} for amorphous and hexagonal ices and $0.125$ g/cm$^3$ for water vapor \cite{GU2022109961}). Note that although the SCS is typically defined as the ESP divided by the molecular density, the molecular and mass densities are related by a constant factor, so comparisons between the different phases are not affected by this convention.

\begin{figure}[t]
    \centering    \includegraphics[width=0.48\textwidth]{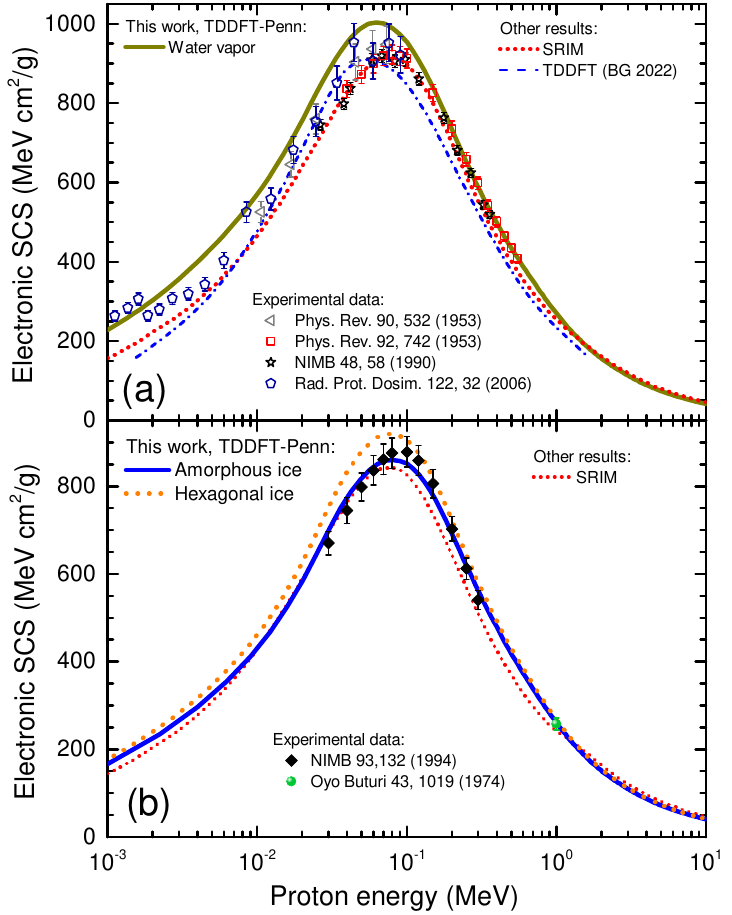}
    \caption{SCS of water vapor, amorphous and hexagonal ice for protons. (a) TDDFT-Penn result (solid black line) is compared with TDDFT by Gu {\it et al.} (dash-dotted green line) \cite{GU2022109961}. The 
    SRIM-2013 (dotted red line) is also presented. Experimental data \cite{PhysRev.90.532,PhysRev.92.742,MITTERSCHIFFTHALER199058,baek2006} are shown by symbols as detailed in the legend.
    (b) TDDFT-Penn results (solid blue line for amorphous ice, dotted olive line for hexagonal ice) are compared with SRIM-2013 
    (dotted red line). Experimental results from Bauer {\it et al.} (black diamonds) \cite{BAUER1994132} and Kamitsubo {\it et al.} (magenta circle) \cite{kamitsubo:1974} are shown for amorphous ice. 
    }
    \label{fig:ice-vapor}
\end{figure}

Figure~\ref{fig:ice-vapor} shows our calculated SCS of water vapor and ices targets for which there is sufficient experimental data to compare with.  
Figure~\ref{fig:ice-vapor}(a) refers to water vapor, and contains our present calculations, the experimental data   \cite{PhysRev.90.532,PhysRev.92.742,MITTERSCHIFFTHALER199058,baek2006}, the results of atomistic real-time TDDFT calculations \cite{GU2022109961}, as well as of the 
SRIM semiempirical curve \cite{srim2010}. 
While the maximum of the ESP calculated by Gu {\it et al.} \cite{GU2022109961} is shifted towards lower proton energies and underestimates the higher energy experimental data, our calculations agree rather well with 
SRIM data \cite{srim2010} and experiments above $\sim 0.2$ MeV, and reasonably well (better than the other reference calculations) with the most recent experimental data \cite{baek2006} around and below the ESP maximum. Figure~\ref{fig:ice-vapor}(b), corresponding to ice, contains only experimental data for the amorphous 
phase \cite{BAUER1994132,kamitsubo:1974}. 
Our results for amorphous and hexagonal ice show a visible phase effect between these two polymorphs, with the former excellently reproducing the experimental data \cite{BAUER1994132,kamitsubo:1974}. Unfortunately, there is no experimental information available for hexagonal ice in the literature. The depicted semiempirical SRIM curve \cite{srim2010} shows appreciable differences around and above the ESP maximum.

Clearly, the TDDFT-Penn method is capable of precisely reproducing the different experimental ESP values in the vapor and ice phases of water in the entire energy range covered by the experiments. In addition, it captures the phase differences, both for the two solid phases studied and also between vapor and ices. 
Therefore, the method is suitable for providing reliable ESP values of liquid water for protons in a wide energy range, as presented in what follows.

\begin{figure}[t]
\centering
\includegraphics[width=0.48\textwidth]{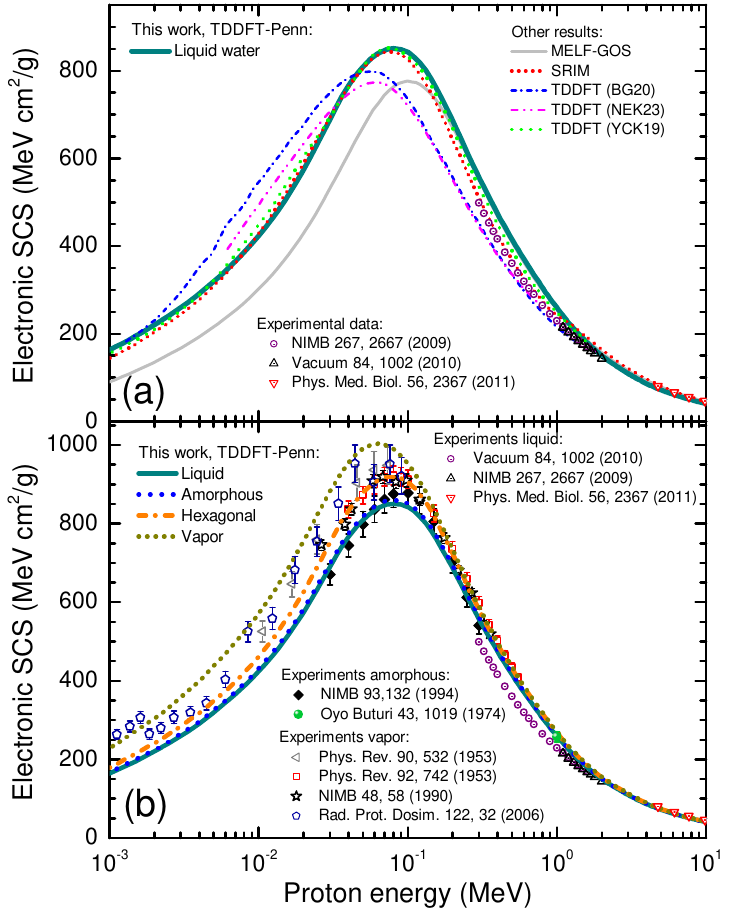}
\caption{(a) SCS of liquid water for protons. TDDFT-Penn (black solid line) is compared with 
SRIM (dotted red line) \cite{srim2010}.
TDDFT are included: dash-dotted wine line \cite{gu2020}, dash-double-dotted olive line \cite{koval2023}, and dash-dotted magenta line \cite{kanai:2019}. MELF-GOS is presented as grey line \cite{GARCIAMOLINA20092647,GarciaMolina2012SpringerScience}. Experimental data include  Shimizu \emph{et al.} (black triangles) \cite{SHIMIZU20092667}, Shimizu \emph{et al.} (purple circles) \cite{SHIMIZU20101002}, and Siiskonen \emph{et al.} (inverted green triangles) \cite{Siiskonen2011}. 
(b) SCS for protons in various water phases normalized by atomic density. 
The symbols represent the experimental data shown in Figs. \ref{fig:ice-vapor} and \ref{fig:lw}(a).}
\label{fig:lw}
\end{figure}

The SCS of liquid water is shown in Figure~\ref{fig:lw}(a), where our real-time TDDFT-Penn calculation is compared to experimental data \cite{SHIMIZU20092667,SHIMIZU20101002,Siiskonen2011,PaulDatabase}
and reference SRIM data 
\cite{srim2010},
as well as the predictions of the MELF-GOS method \cite{GARCIAMOLINA20092647,GarciaMolina2012SpringerScience} (based on the dielectric response formalism) and of atomistic real-time TDDFT \cite{gu2020,koval2023,kanai:2019}. 
For proton energies $E>0.2$ MeV, good agreement is observed between TDDFT-Penn and MELF-GOS \cite{GARCIAMOLINA20092647,GarciaMolina2012SpringerScience}.
Two real-time TDDFT atomistic results 
\cite{gu2020,koval2023} 
agree with each other for energies $E > 0.1$ MeV but show significant differences below this energy, and predict the ESP maximum at lower energies than other calculations. They also underestimate our present results, as well as MELF-GOS calculations and 
SRIM data for energies around and above the ESP maximum.
The TDDFT-Penn results show excellent agreement with those of atomistic calculations by Yao {\it et al.} \cite{kanai:2019}.
Both semiempirical SRIM values \cite{srim2010} and the experimentally-derived data 
\cite{SHIMIZU20092667,SHIMIZU20101002} clearly fall below our present and MELF-GOS results, what motivates a more in-depth discussion in what follows.




Experimental SCS data for liquid water
\cite{SHIMIZU20092667,SHIMIZU20101002} were derived from transmission proton energy loss through cylindrical liquid water jets, by modifying the SRIM SCS values \cite{srim2010} until the Monte Carlo simulated spectra and measurements coincided.
This procedure was critically analyzed \cite{GARCIAMOLINA201351,deVera2019jets}, after perfectly reproducing experimental energy spectra by the SEICS code \cite{GarciaMolina2011} using the SCS values obtained by the MELF-GOS method, only assuming a slight reduction in the liquid jet diameter (due to water evaporation). The excellent agreement between the present TDDFT-Penn and MELF-GOS results in the $0.2$-$2$ MeV range supports the accuracy of our theoretical values in this energy range and the suggested underestimation \cite{GARCIAMOLINA201351,deVera2019jets} of the experimental SCS values in this energy range \cite{SHIMIZU20092667,SHIMIZU20101002}. There is excellent agreement between the experimental results above $4$ MeV \cite{Siiskonen2011}, obtained by direct transmission measurements in liquid water films, and our calculations.


The TDDFT-Penn and experimental SCS for the different water phases are gathered in Figure~\ref{fig:lw}(b). 
Sizeable phase differences can be clearly observed for proton energies around and below the maximum, where vapor SCS is notably larger than that of the condensed phases, due to the lower electron screening ($N_{\rm e}^{\rm vapor} = 2 > N_{\rm e}^{\rm cond.\, phases} \simeq 1$). The SCS for hexagonal ice and water vapor are larger than for liquid water and amorphous ice at energies around the maximum, whereas the SCS for all phases converge above $\sim 0.5$ MeV. Of particular relevance is the fact that the curves for liquid water and amorphous ice are practically identical in the whole energy range. All the features displayed by our TDDFT-Penn calculations are also seen in the experimental SCS for the different phases of water.

The observed phase effects for the energy loss of protons in water are of relevance
in proton therapy, where the precise position of the Bragg peak depends on the details of the ESP curve.
Notably, the practical coincidence between the SCS of liquid water and amorphous ice for proton beams is 
consistent with recent findings for electron projectiles \cite{Signorell2020}. 
This similarity provides valuable information, allowing researchers to obtain otherwise elusive energy loss quantities for charged particles in the biologically-relevant liquid water by using data from amorphous ice. Moreover, our results for different phases of water are important in the context of the interaction of cosmic rays with water vapor and ice in atmospheric and space environments, where accurate knowledge of proton energy loss is essential for modeling radiation effects. In particular, understanding how phase transitions influence stopping power contributes to more reliable simulations of astrochemical processes and the radiolytic evolution of ices on planetary surfaces, comets, and interstellar dust grains.
We thank financial support by:
IPEN Proj. No. 2020.06.IPEN.32, CNPq Proj. No. 406982/2021-0 and 403722/2023-3, Coordenação de Aperfeiçoamento de Pessoal de Nível Superior, Finance Code 001, FINEP, Instituto Nacional de Engenharia de Superfícies, PRONEX-FAPERGS, Spanish Ministerio de Ciencia e Innovación Proj. PID2021-122866NB-I00 financed by MCIN/AEI/10.13039/501100011033/ and ERDF A way
of making Europe, Fundación Séneca – Agencia de Ciencia y Tecnología de la Región de Murcia Proj. 22081/PI/22, FAPESP computer cluster process Nos. 2012/04583-8 and 2020/04867-2. T.F.S. and M.H.T. acknowledges the support provided by CNPq-INCT-FNA Project No. 464898/2014-5. N.E.K. acknowledges the funding by the Basque Government Education Departments’ IKUR program co-funded by the European NextGenerationEU action through the Spanish Plan de Recuperación, Transformación y Resiliencia (PRTR).

\bibliography{manuscript-resubmit}

\end{document}



\title{%
  \begin{flushleft}
    Supplemental material for:
  \end{flushleft}
  \centering
  Stopping cross-section for protons across different phases of water
}

\author{F. Matias\textsuperscript{1}}
\author{N. E. Koval\textsuperscript{2}}
\author{P. de Vera\textsuperscript{3}}
\author{R. Garcia-Molina\textsuperscript{3}}
\author{I. Abril\textsuperscript{4}}
\author{J. M. B. Shorto\textsuperscript{1}}
\author{H. Yoriyaz\textsuperscript{1}}
\author{J. J. N. Pereira\textsuperscript{1}}
\author{T. F. Silva\textsuperscript{5}}
\author{M. H. Tabacniks\textsuperscript{5}}
\author{M. Vos\textsuperscript{6}}


\author{P. L. Grande\textsuperscript{7}}

\affiliation{\textsuperscript{1}Instituto de Pesquisas Energéticas e Nucleares, Av. Professor Lineu Prestes, São Paulo, 05508-000, Brazil}

\affiliation{\textsuperscript{2}Centro de F\'isica de Materiales, Paseo Manuel de Lardiz\'abal 5, Donostia-San Sebasti\'an, 20018, Spain}

\affiliation{\textsuperscript{3}Departamento de Física, Centro de Investigación en Óptica y Nanofísica, Universidad de Murcia, Murcia, Spain}

\affiliation{\textsuperscript{4}Departament de Física Aplicada, Universitat d’Alacant, Alacant, Spain}

\affiliation{\textsuperscript{5}Instituto de Física da Universidade de São Paulo, Rua do Matão, trav. R187, São Paulo, 05508-090, Brazil}

\affiliation{\textsuperscript{6}Department of Materials Physics, Research School of Physics, Australian National University (ANU), Canberra, Australia}

\affiliation{\textsuperscript{7}Instituto de Física da Universidade Federal do Rio Grande do Sul, Av. Bento Gonçalves, Porto Alegre, 9500, Brazil}

\date{\today}


\maketitle


\section*{\NoCaseChange{Details on the TDDFT-Penn methodology to calculate the electronic stopping power for protons}}
\label{sec:theory}

In this work we combine real-time TDDFT \cite{borisov2004,quijada2007,koval2013,matias2017} and the Penn model \cite{penn1987,Maarten:2019b}, a method which we 
refer to as the TDDFT-Penn approach \cite{matias:2024}. Instead of simulating an explicit atomistic structure, the interaction of an energetic proton with a nearly homogeneous electron gas (jellium) of a given density is efficiently treated by real-time TDDFT. The water media are modeled as jellium spheres in all the DFT and real-time TDDFT calculations. 

The positive background density of the jellium with radius $R_{\rm cl}$ is defined by $n_{0}^{+}({\bf r})=n_{0}^{+}(r_s) \Theta (R_{\rm cl}-r)$, where $\Theta(x)$ denotes the Heaviside step-function and $n_{0}^{+}(r_s)$ is the constant bulk density, which depends only on the Wigner-Seitz radius $r_{s}$: ($4\pi r_{s}^{3}/3)=1/n_{0}$. The total number of electrons in the neutral clusters, $N_{e}$, is then given by $N_{e}=(R_{\text{cl}}/r_{s})^{3}$. Thus, the size of each closed-shell cluster is determined by the density parameters $r_s$ and the total number of electrons, $N_{e}=588$. According to the ELFs of the water phases (see Fig.~1 in the main text), and using the relation $\omega_{\rm p}^2=4\pi n_0$, the most important Wigner-Seitz radii vary from $r_s = 0.6$ ($\sim 100$ eV) to $5.0$ ($\sim 4.2$ eV) au. The jellium spheres corresponding to this range have sizes varying from $R_{\text{cl}} \sim 5.0$ to $\sim 42.0$ au.

A static density functional theory (DFT) calculation is performed to obtain the system's ground state. The time evolution of the complete electronic density, $n({\bf r},t)$, in response to an external field (in this case, a proton), is conducted within the framework of real-time TDDFT in the Kohn-Sham regime (atomic units are used throughout unless specified otherwise):
\begin{equation}
    i\frac{\partial \psi _{j}({\bf r},t)}{\partial t} ~=~\left\{
T+V_{\rm{eff}}([n],{\bf r},t)\right\} \psi _{j}({\bf
r},t)~\mathrm{,} \label{kseq}
\end{equation}

\noindent
where $\psi_j({\bf r},t)$ are the Kohn-Sham orbitals and $T$ is the kinetic energy operator. The Kohn-Sham effective potential, $V_{\text{eff}}([n],{\bf r},t)$, is a function of the electronic density of the system:
$n({\bf r},t)= \sum_{j\in {\rm occ.}}{\left| \psi_{j}({\bf r},t)\right|
^{2}}$. 

The effective potential $V_{\text{eff}}=V_{\rm{ext}}^+({\bf r})+V_{\rm{H}}([n],{\bf r},t)+V_{\rm{xc}}([n],{\bf r},t)+V_{\rm{p}}({\bf r},t)$ is obtained as the sum of the external potential created by the positive background of the jellium sphere $V_{\text{ext}}^+({\bf r})$, the Hartree potential $V_{\text{H}}([n],{\bf r},t)$, the exchange-correlation potential $V_{\text{xc}}([n],{\bf r},t)$, and the potential representing the projectile $V_{\rm{p}}({\bf r},t)$, which is modeled as a bare Coulomb charge. The $V_{\text{xc}}([n],{\bf r},t)$ is calculated using the Gunnarsson and Lundqvist kernel \cite{PhysRevB.13.4274} within the standard adiabatic local density approximation (ALDA) approach. In these simulations, the proton traverses the target through the geometric center. 
This methodology has been successfully applied to calculate the stopping power of protons in various metallic and nonmetallic targets \cite{PhysRevLett.81.4831,quijada2007,Koval2017}.

The ESP is calculated by integrating the time-dependent induced force over the whole proton trajectory:
\begin{equation}
E_{\rm{loss}}(v,\omega_{\rm p})=-v\int_{-\infty}^{+\infty} F_z (t){\rm d}t,
\end{equation}
\noindent
where $v$ is the (constant) velocity at which the proton traverses the jellium and $\omega_{\rm p}$ is the frequency of the plasmon, a value determined by the individual contributions of the electron gas obtained from $r_s$; $\omega_{\rm p}=\sqrt{3}r_s^{-3/2}$. Once the induced force on the proton is calculated, the average or effective ESP is computed as the energy loss per unit path length, i.e.,
\begin{equation}
\left[\frac{{\rm d}E}{{\rm d}z}(v,\omega_{\rm p})\right]_{\text{TDDFT}}=\frac{E_{\rm{loss}}(v,\omega_{\rm p})}{2R_{\text{cl}}}.
\label{stp_tddft}
\end{equation}

Then, the Penn model is used to represent the inhomogeneities of the target's electron density by means of its ELF. As previously shown for such complex targets as polymers \cite{matias:2024}, the non-perturbative character of the TDDFT method allows the precise calculation of proton ESP at energies even below the ESP maximum, while the Penn approach realistically captures the particular electronic structure of the complex material. 

To achieve this goal, each free electron density is analyzed based on the material's ELF at the optical limit, as follows \cite{Maarten:2019b}:
\begin{eqnarray}
       g(\omega_{\rm p}) =\frac{2}{\pi \omega_{\rm p}}\text{ELF}(\omega_{\rm p}).
    \label{eq:elf} 
\end{eqnarray}
\noindent
The ESP depends on the frequency of the plasmon, $\omega_{\rm p}$. Therefore, the ESP is now calculated within the real-time TDDFT-Penn approach as follows \cite{matias:2024}:
\begin{equation}
    \left[\frac{{\rm d}E}{{\rm d}z}(v)\right]_{\text{TDDFT-Penn}} = \int_0^{\infty}{\rm d}\omega_{\rm p} g(\omega_{\rm p})\left[\frac{{\rm d}E}{{\rm d}z}(v,\omega_{\rm p})\right]_{\text{TDDFT}}.
    \label{eq:penn}
\end{equation}
\noindent
In the above equation, the term $\left[{\rm d}E/{\rm d}z(v,\omega_{\rm p})\right]_{\text{TDDFT}}$ is calculated in the real-time TDDFT framework using Equation (\ref{stp_tddft}). 





\newpage

\bibliography{supplemental-material}